\documentclass[aps,prc,twocolumn,showpacs,preprintnumbers,amsmath,amssymb,superscriptaddress]{revtex4-1}

\usepackage{graphicx}
\usepackage{epsfig}
\usepackage{dcolumn}
\usepackage{bm}
\usepackage{epstopdf}
\usepackage{float}
\usepackage{titlesec}

\titlespacing*{\section}
{0pt}{12pt plus 2pt minus 2pt}{12pt plus 2pt minus 2pt}

\begin{document}


\title{Nuclear level densities and gamma-ray strength functions of $^{180,181,182}$Ta}

\author{C.P.~Brits}
\affiliation{Department of Subatomic Physics, iThemba LABS, P.O. Box 722, Somerset West 7129, South Africa}
\affiliation{Physics Department, Stellenbosch University, Matieland 7602, South Africa}
\author{K.L.~Malatji}
\email{klmalatji@tlabs.ac.za}
\affiliation{Department of Subatomic Physics, iThemba LABS, P.O. Box 722, Somerset West 7129, South Africa}
\affiliation{Physics Department, Stellenbosch University, Matieland 7602, South Africa}
\affiliation{Physics Department, University of Western Cape, Bellville 7535, South Africa}
\author{M. Wiedeking}
\email{wiedeking@tlabs.ac.za}
\affiliation{Department of Subatomic Physics, iThemba LABS, P.O. Box 722, Somerset West 7129, South Africa}
\author{B.V.~Kheswa}
\affiliation{Department of Subatomic Physics, iThemba LABS, P.O. Box 722, Somerset West 7129, South Africa}
\affiliation{Department of Applied Physics and Engineering Mathematics, University of Johannesburg, Doornfontein 2028, South Africa}
\author{S.~Goriely}
\affiliation{Institut d'Astronomie et d'Astrophysique, Universit\'e Libre de Bruxelles, CP 226, B-1050 Brussels, Belgium}
\author{F.L.~Bello Garrote}
\affiliation{Department of Physics, University of Oslo, N-0316, Oslo, Norway}
\author{D.L.~Bleuel}
\affiliation{Lawrence Livermore National Laboratory, Livermore, CA 94550-9234, USA}
\author{F.~Giacoppo}
\affiliation{Department of Physics, University of Oslo, N-0316, Oslo, Norway}
\affiliation{Helmholtz Institute Mainz, 55099 Mainz, Germany}
\affiliation{GSI Helmholtzzentrum f{\"u}r Schwerionenforschung, 64291 Darmstadt, Germany}
\author{A.~G{\"o}rgen}
\affiliation{Department of Physics, University of Oslo, N-0316, Oslo, Norway}
\author{M.~Guttormsen}
\affiliation{Department of Physics, University of Oslo, N-0316, Oslo, Norway}
\author{K.~Hadynska-Klek}
\affiliation{Department of Physics, University of Oslo, N-0316, Oslo, Norway}
\author{T.W.~Hagen}
\affiliation{Department of Physics, University of Oslo, N-0316, Oslo, Norway}
\author{S. Hilaire}
\affiliation{CEA, DAM, DIF, F-91297 Arpajon, France}
\author{V.W.~Ingeberg}
\affiliation{Department of Physics, University of Oslo, N-0316, Oslo, Norway}
\author{H.~Jia}
\affiliation{Institute of Space Sciences, Shandong University, Weihai 263209, China}
\author{M.~Klintefjord}
\affiliation{Department of Physics, University of Oslo, N-0316, Oslo, Norway}
\author{A.C.~Larsen}
\affiliation{Department of Physics, University of Oslo, N-0316, Oslo, Norway}
\author{S.N.T.~Majola}
\affiliation{Department of Subatomic Physics, iThemba LABS, P.O. Box 722, Somerset West 7129, South Africa}
\affiliation{Physics Department, Stellenbosch University, Matieland 7602, South Africa}
\author{P.~Papka}
\affiliation{Department of Subatomic Physics, iThemba LABS, P.O. Box 722, Somerset West 7129, South Africa}
\affiliation{Physics Department, Stellenbosch University, Matieland 7602, South Africa}
\author{S. P\'eru}
\affiliation{CEA, DAM, DIF, F-91297 Arpajon, France}
\author{B.~Qi}
\affiliation{Institute of Space Sciences, Shandong University, Weihai 263209, China}
\author{T.~Renstr{\o}m}
\affiliation{Department of Physics, University of Oslo, N-0316, Oslo, Norway}
\author{S.J.~Rose}
\affiliation{Department of Physics, University of Oslo, N-0316, Oslo, Norway}
\author{E.~Sahin}
\affiliation{Department of Physics, University of Oslo, N-0316, Oslo, Norway}
\author{S.~Siem}
\affiliation{Department of Physics, University of Oslo, N-0316, Oslo, Norway}
\author{G.M.~Tveten}
\affiliation{Department of Physics, University of Oslo, N-0316, Oslo, Norway}
\author{F.~Zeiser}
\affiliation{Department of Physics, University of Oslo, N-0316, Oslo, Norway}

\date{\today}
 
\begin{abstract}
\noindent{}Particle-$\gamma$ coincidence experiments were performed at the Oslo Cyclotron Laboratory with the $^{181}$Ta(d,X) and $^{181}$Ta($^{3}$He,X) reactions, to measure the nuclear level densities (NLDs) and $\gamma$-ray strength functions ($\gamma$SFs) of $^{180, 181, 182}$Ta using the Oslo method. The Back-shifted Fermi-Gas, Constant Temperature plus Fermi Gas, and Hartree-Fock-Bogoliubov plus Combinatorial models where used for the absolute normalisations of the experimental NLDs at the neutron separation energies. The NLDs and $\gamma$SFs are used to calculate the corresponding $^{181}$Ta(n,$\gamma$) cross sections and these are compared to results from other techniques. The energy region of the scissors resonance strength is investigated and from the data and comparison to prior work it is concluded that the scissors strength splits into two distinct parts. This splitting may allow for the determination of triaxiality and a $\gamma$ deformation of $14.9^{\circ} \pm 1.8^{\circ}$ was determined for $^{181}$Ta. 
\end{abstract}


\maketitle

\section{\label{sec:level1}Introduction}

\noindent{}The $\gamma$-ray strength function ($\gamma$SF) and nuclear level density (NLD) describe the nuclear structure in the region of the quasi-continuum where the level spacing is too small to resolve and study individual levels. The $\gamma$SF characterises the average electromagnetic properties and is related to radiative decay and photo-absorption processes \cite{Chadwick2011, Bar1973}. From the NLD the evolution of the number of levels with excitation energy can be investigated \cite{Gut2015} and related to thermodynamic properties \cite{Mor2015}. 

The $\gamma$SF and NLD are important input parameters into reaction cross section calculations in the Hauser-Feshbach statistical framework \cite{Hauser1952}. The Hauser-Feshbach formalism is implemented in the TALYS v1.9 reaction code \cite{Koning2008} which can be used to calculate (n,$\gamma$) cross sections. Hence, NLD and $\gamma$SF are nuclear properties of significance to nucleosynthesis \cite{Arnould2007} and calculations have shown that relative small changes to the overall shape of the $\gamma$SF, such as a pygmy resonance, can have an order of magnitude effect on the rate of elemental formation \cite{Goriely1998}. It has been shown that measured statistical properties can reliably be used to reproduce capture cross sections that were measured using other techniques \cite{Guttormsen2017,Kheswa2017,Larsen2016}, although further validations are needed across the nuclear chart. Additionally, NLD and $\gamma$SF can also be relevant to the design of existing and future nuclear power reactors, where simulations depend on such nuclear data \cite{Chadwick2011}. Their importance is highlighted by the efforts which are currently underway to generate a reference database for $\gamma$SFs \cite{Vivian2018}. 

A key feature of the $\gamma$SF in well-deformed nuclei is the scissors resonance (SR). The SR is a collective magnetic dipole (M1) excitation usually found at excitation energies $E_{x} \approx$ 2-4 MeV. The SR was predicted several decades ago \cite{Suz1977, Iudice1978, Iud1979, Iac1981} and first observed in $^{156}$Gd a few years later \cite{Bohle1984}. A splitting of the SR in $^{164}$Dy and $^{174}$Yb was reported soon after \cite{Boh1984} and interpreted as a possible measure of nuclear triaxiality \cite{Iudice1985}. Besides observations in well-deformed even-even nuclei (\cite{Kris2010} and references therein), the SR has also been observed in less deformed nuclei, e.g. in vibrational even-mass $^{122-130}$Te \cite{Schwengner1997}, transitional $^{190,192}$Os \cite{Fransen1999}, and in $\gamma$-soft $^{134}$Ba and $^{196}$Pt \cite{Maser1996, Brentano1996} nuclei. The SR has been investigated through nuclear resonance fluorescence (NRF) \cite{Kneissl1996}, resonance neutron capture \cite{Bar2015}, and through the Oslo Method in the rare-earth \cite{Schiller2000} and actinide \cite{Guttormsen2012, Guttormsen2014, Tornyi2014, Laplace2016} regions. A review of the theoretical and experimental findings can be found in Ref. \cite{Kris2010}.

In this paper we present results of the NLDs and $\gamma$SFs for $^{180,181,182}$Ta from six reactions. Three different level density models are used and compared for the normalisation at $S_{n}$. From the (d,p)$^{182}$Ta data the $^{181}$Ta(n,$\gamma$) cross section is calculated using TALYS and compared to previous results. The emergence of the SR in the transitional nucleus $^{181}$Ta is investigated and compared to other work. The paper is structured as follows: in Sec. II the experimental setup is presented and Sec. III provides a brief overview of the Oslo method and the different level density models that were used. Sec. IV presents the $^{181}$Ta(n,$\gamma$) cross section and a comparison to other work, while Sec. V investigates and discusses the presence of the SR in $^{181}$Ta. A brief summary is given in Sec. VI.

\section{Experimental Setup}
\label{sec:level3}

\noindent{}Three experiments were performed at the Oslo Cyclotron Laboratory (OCL) at the University of Oslo using a self-supporting 0.8 mg/cm$^{2}$ thick natural tantalum target. A deuteron beam of 12.5 MeV was used for the $^{181}$Ta(d,p)$^{182}$Ta and $^{181}$Ta(d,d')$^{181}$Ta reactions, while a deuteron beam of 15 MeV was used for the $^{181}$Ta(d,t)$^{180}$Ta reaction and a second $^{181}$Ta(d,d')$^{181}$Ta reaction. A 34 MeV $^{3}$He beam was utilised for the $^{181}$Ta($^{3}$He,$^{3}$He')$^{181}$Ta and $^{181}$Ta($^{3}$He,$\alpha$)$^{180}$Ta reaction. The SiRi particle telescope \cite{Guttormsen2010} and CACTUS scintillator \cite{Guttormsen1990} array were used to detect charged particles and $\gamma$-rays in coincidence within a 2$\mu$s hardware time window. 

The $\Delta$E-E SiRi particle-telescope consists of eight 130 $\mu$m thin, segmented silicon $\Delta$E detectors and eight 1550 $\mu$m thick E silicon detectors. These detectors covered a polar angular range of $\theta_{lab} = 126^{\circ} - 140^{\circ}$ with respect to the beam axis. The energy resolutions, as determined from the elastic peaks, are $\approx$ 125 keV for the deuteron and 350 keV for the $^{3}$He beams. The CACTUS array consists of 26 NaI(Tl) detectors with $5" \times 5"$ crystals positioned 22 cm away from the target, covering a solid angle of 16.2$\%$ of $4\pi$ sr. CACTUS has a total efficiency of 14.1(1)$\%$ and an energy resolution of 6$\%$ FWHM for a 1332 keV $\gamma$-ray transition. 

The E detectors provided the start signal and the delayed NaI(Tl) detectors provided the stop signal for the time-to-digital converters, enabling event-by-event sorting for the particle-$\gamma$ coincidence data. Calibrations of SiRi was accomplished using individual reactions on $^{181}$Ta. CACTUS detectors was calibrated with the $^{28}$Si(d,p) reaction which provided appropriate $\gamma$-ray energies. During offline analysis the prompt time gate was set to 40 ns for the data sets from $^{3}$He beams and to 30 ns for the data from deuteron beams. Equivalently wide non-prompt time gates were used to subtract and remove the uncorrelated events from the prompt particle-$\gamma$ events. 

\section{Analysis}
\subsection{Oslo Method}

\noindent{}The $\gamma$SFs and NLDs are simultaneously extracted using the Oslo Method, which has been covered in the literature \cite{Schiller2000, Larsen2011, Guttormsen1996, Guttormsen1987}, and only a brief overview will be presented here. In the first step the $\gamma$-ray spectra is unfolded using the detector response function. The Compton background, effects from pair production and the single- and double-escape peaks are removed from the $\gamma$-ray spectrum leaving only full-energy deposit events that are corrected for efficiency. The primary $\gamma$-rays are extracted using an iterative subtraction method that separates the primary $\gamma$-rays from the total $\gamma$-ray cascade. The primary transitions are collected in the first-generation matrix $P(E_{x},E_{\gamma})$ with the assumption that the $\gamma$-ray distribution is the same for a state populated through $\gamma$-ray decay or the nuclear reaction. This assumption is valid at high-level densities where the nucleus is in a compound state prior to $\gamma$-ray emission.

The probability for a $\gamma$-ray, with energy $E_{\gamma}$, to decay from excitation energy $E_{x}$ to a final energy $E_{f}$, with energy $E_{f} = E_{x} - E_{\gamma}$, is proportional to the level density at the final energy, $\rho(E_{f})$ and the transmission coefficient $\mathcal{T}(E_{\gamma})$. $P(E_{x},E_{\gamma})$ is proportional to the decay probability and can be factorised as:

\begin{equation}
P(E_x,E_{\gamma}) \propto \mathcal{T}(E_{\gamma})\rho(E_{f}).
\end{equation}

\noindent{}Brink's hypothesis \cite{Bri1955} is assumed to be valid, which implies that the $\gamma$-ray transmission coefficient does not depend on the properties of the initial and final states but only on the $\gamma$-ray energy. A $\chi^{2}$ minimisation is used to extract $\mathcal{T}(E_{\gamma})$ and $\rho(E_{f})$ \cite{Schiller2000}: 

\begin{equation}
\begin{split}
\chi^{2} = \frac{1}{N_{free}} \sum^{E^{max}}_{E_{x}=E^{min}} \sum^{E_{x}}_{E_{\gamma}=E^{min}_{\gamma}} \times \\
\left(\frac{P_{th}(E_{x},E_{\gamma}) - P(E_{x},E_{\gamma})}{\delta P(E_{x},E_{\gamma})}\right)^2,
\end{split}
\end{equation}

\begin{figure}[h!]
	\centering
	\includegraphics[width=0.5\textwidth]{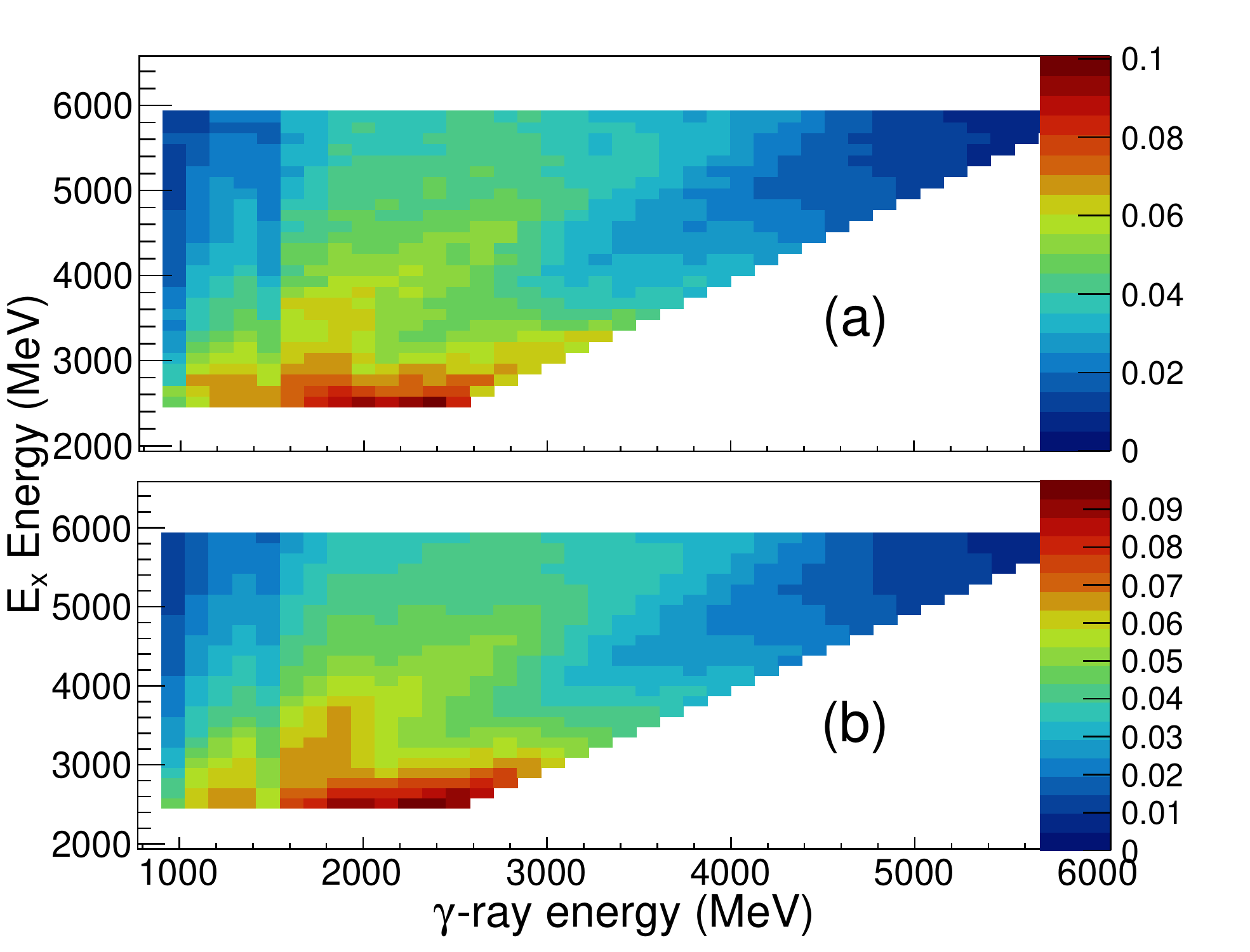}
	\caption{(Color online) The experimental (a) and fitted (b) first-generation particle-$\gamma$ matrices from the $^{181}$Ta(d,p)$^{182}$Ta reaction with a deuteron energy of 12.5 MeV.} 
	\label{rhosigchi}
\end{figure}

\noindent{}where $N_{free}$ is the number of degrees of freedom and $\delta P(E_{x},E_{\gamma})$ is the uncertainty in the first-generation matrix. The experimental $P(E_{x},E_{\gamma})$ and fitted $P_{th}(E_{x},E_{\gamma})$ first-generation matrices for $^{182}$Ta are shown in Fig. \ref{rhosigchi}. Their close similarity encourages an accurate fit. The $\chi^{2}$ minimisation was applied in the regions shown in Tab. \ref{one_two_three}.

\begin{table}[!hbt]
\centering
\caption{The regions where the $\chi^{2}$ minimisation was applied to data from the different reactions populating $^{180,181,182}$Ta.}
\setlength{\tabcolsep}{6.2pt}
\begin{tabular}{|c|c|c|c|c|}
\hline
Reaction & E$_{beam}$ & E$_{\gamma}^{min}$& E$_{x}^{min}$ & E$_{x}^{max}$\\
~&(MeV)&(MeV)&(MeV)&(MeV)\\

\hline
($^{3}$He,$\alpha$)$^{180}$Ta&34&1.73 & 2.97 & 6.35\\
($^{3}$He,$^{3}$He')$^{181}$Ta&34& 1.63 & 2.57 & 7.38\\
(d,t)$^{180}$Ta&15& 1.21 & 2.49 & 5.18\\
(d,d')$^{181}$Ta&15& 1.21 & 3.01 & 6.02\\
(d,d')$^{181}$Ta&12.5& 1.59 & 2.54 & 3.84\\
(d,p)$^{182}$Ta&12.5& 1.54 & 2.54 & 5.94\\
\hline
\end{tabular} \label{one_two_three}
\end{table}

Within these limits an infinite number of solutions for $P(E,E_{\gamma})$ can be found of the form:

\begin{equation}
\label{level}
\tilde{\rho}(E_{f}) = Ae^{\alpha E_{f}}\rho(E_{f}) 
\end{equation}
and 
\begin{equation}
\label{transmission}
\widetilde{\mathcal{T}}(E_{\gamma}) = Be^{\alpha E_{\gamma}}\mathcal{T}(E_{\gamma}),
\end{equation}

\noindent{}where $A$ and $B$ are normalisation parameters and $\alpha$ is the slope of the NLD and $\gamma$-ray transmission coefficient.

\subsection{Nuclear level density}\label{NLD_label}

\noindent{}A normalisation is performed to determine the parameters $A$ and $B$ and the slope $\alpha$, corresponding to the physical solutions, from other experimental data as well as systematics. The NLD is normalised at low energies to experimentally measured levels by counting the levels from the evaluated nuclear data base \cite{nndc}. At high $E_{x}$ the NLD is normalised to the total level density at the neutron separation energy $\rho(S_{n})$. 

The functional form of the NLD is uniquely defined from the $\chi^{2}$ fit of the primary $\gamma$-ray matrix. It is for the absolute normalisation at the neutron separation energy that different level density models, in particular the spin distribution, play a major role. For this work three different normalisation models are considered. The Back-shifted Fermi-Gas (BSFG) \cite{Gilbert1965}, Constant Temperature+Fermi Gas (CT+FG) \cite{Koning2012}, and Hartree-Fock-Bogoliubov plus Combinatorial (HFB) \cite{Goriely2008}.

The CT+FG normalisation is based on two different spin cut-off formulas. Firstly, using the energy-dependent spin cut-off parameter, the NLD can accurately be obtained from the widely used Constant Temperature model (CT) \cite{Gilbert1965}, for 2$\Delta_{0}\leq E_{x}\leq$10 MeV, where $\Delta_{0}$ is the pair-gap parameter \cite{Bohr1969}. The total NLD $\rho(S_{n})$ is calculated according to \cite{Larsen2011}:

\begin{equation}\label{NLD_rho}
\rho(S_{n}) = \frac{2\sigma^{2}}{D_{0}} \times \frac{1}{(I+1)\text{exp}(-\frac{(I+1)^{2}}{2\sigma^{2}}) + I\text{exp}(-\frac{I^{2}}{2\sigma^{2}})}.
\end{equation}

\noindent{}$D_{0}$ is the $\ell = 0$ neutron resonance spacing data \cite{Capote, Mughabghab2006}, $I$ is the initial spin of the target nucleus, and the spin cut-off parameter $\sigma$ is determined from \cite{Egiby2009}:

\begin{equation}\label{spin_cut_eqn}
\sigma^{2} = 0.391A^{0.675}(E_{x}-0.5Pa)^{0.312}
\end{equation}

\begin{figure}[h!]
	\includegraphics[width=0.53\textwidth]{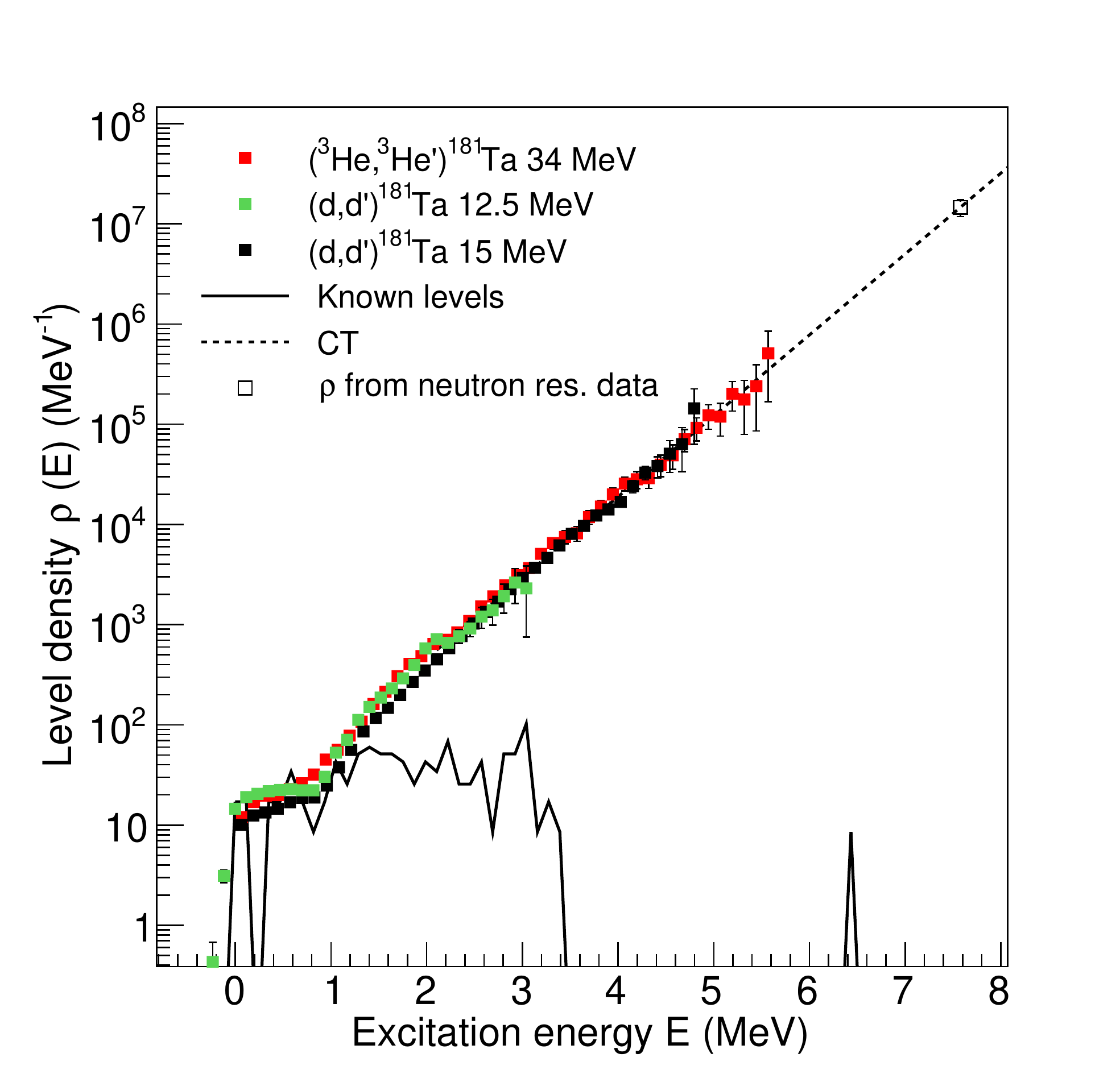}
	\caption{(Color online) The NLDs of $^{181}$Ta from the 12.5 MeV $^{181}$Ta(d,d') (green), 15 MeV $^{181}$Ta(d,d') (black) and $^{181}$Ta($^{3}$He,$^{3}$He') (red) reactions using the CT+FG1 model. The solid line represents the level density deduced from known levels. The dashed line from the CT model \cite{Koning2008b}, interpolates between the experimental data and $\rho(S_{n})$ (black open square).}
	\label{NLD_3}
\end{figure}

\noindent{}where $A$ is the number of nucleons and $Pa$ is the deuteron pairing energy. When using this spin distribution the model will be referred to as CT+FG1. Since the NLD can only be extracted up to $E_x-E_{\gamma}$ and does not reach $S_{n}$, the CT model \cite{Koning2008b} is used to interpolate between the experimental NLD and $\rho(S_{n})$. The experimentally extracted $^{181}$Ta NLD with CT+FG1 from all three reactions populating $^{181}$Ta are shown in Fig. \ref{NLD_3} and are in good agreement. Secondly, the CT+FG normalisation uses the spin cut-off parameter as implemented in TALYS \cite{Koning2008}. The $E_{x}$ is divided into two excitation energy regions: 0 $\leq$$E_{x}$ $\leq$ $E_{M}$, where the constant temperature approximation applies and $E_{x} > E_{M}$, where the Fermi-gas model applies \cite{Ericson1959}. $E_{M}$ is the matching excitation energy between the two models. When using the spin distribution from TALYS the model will be referred to as CT+FG2.

The microscopic HFB model describes the energy-, spin- and parity-dependent NLD. This model takes into account the HFB single-particle level scheme to calculate incoherent intrinsic state densities which depends only on $E_{x}$, parity and the spin projection on the symmetry axis of the nucleus. The collective (rotational and vibrational) enhancement are accounted for, once the incoherent particle-hole states densities have been determined. The resulting microscopic approach reproduces well the experimental data at known discrete states and $S_{n}$. These NLDs are tabulated in the TALYS software package. 

The BSFG model \cite{Gilbert1965,Dilg1973} for the NLD is based on the Fermi-gas approximation and includes pairing energies and shell correction effects in its calculations. In this model the level density parameter and energy shift are free parameters to allow for a reasonable fit to experimental data. 

In the case of $^{180}$Ta, neither $D_{0}$ nor the average radiative width, $\langle \mathit{\Gamma_{\gamma 0}} \rangle$ are known. The $\rho(S_{n})$ was estimated by normalising both $\rho(E_{x})$ and $\mathcal{T}(E_{\gamma})$ of $^{180}$Ta on the basis of these functions having the same slope as $\rho(E_{x})$ and $\mathcal{T}(E_{\gamma})$ of $^{181,182}$Ta using eqn. \ref{NLD_rho}. It has been shown that $\rho(E_{x})$ and $\mathcal{T}(E_\gamma)$ of neighbouring isotopes have the same slope \cite{Mor2015}, independent of the normalisation method used. The spline fit function, as implemented in TALYS \cite{Koning2008}, was used to estimate $\langle \mathit{\Gamma_{\gamma 0}} \rangle$. 

\begin{figure}[h!]
	\centering
	\includegraphics[width=0.53\textwidth]{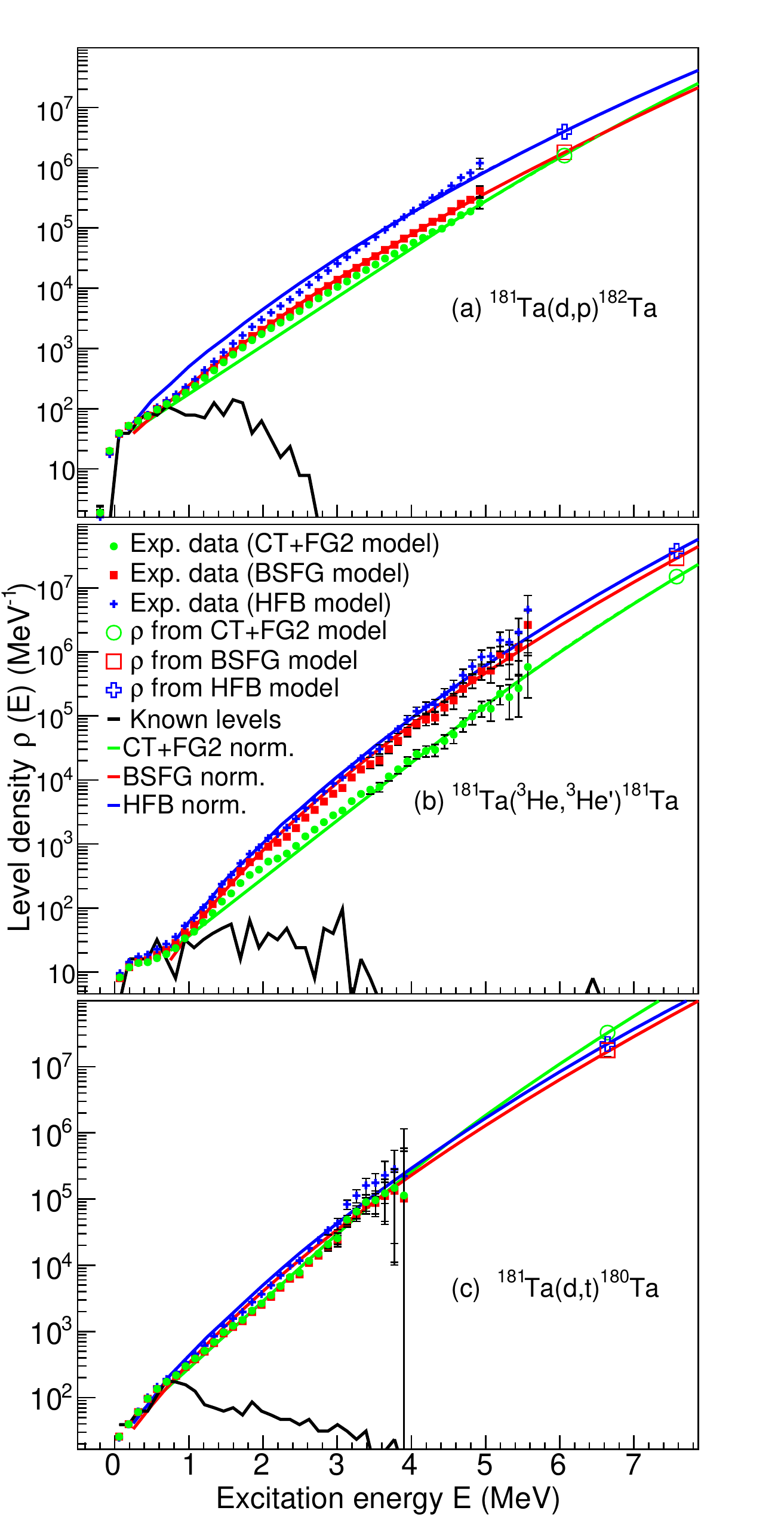}
	\caption{(Color online) The NLD of $^{182}$Ta from the (d,p) reaction (a), $^{181}$Ta from the ($^{3}$He,$^{3}$He') reaction (b), and $^{180}$Ta from the (d,t) reaction (c) are shown with CT+FG2, BSFG and HFB normalisations (see text for details).}
	\label{182Ta_NLD_Model}
\end{figure}

The NLDs of $^{180,181,182}$Ta using the three normalisations are shown in Fig. \ref{182Ta_NLD_Model}. The open squares are the $\rho(S_{n})$ and the solid lines are the level density calculated by the individual models. The experimental data are then normalised to these calculations and are superimposed for comparison. All the models reproduced the $D_{0}$ within experimental uncertainties. The different models will be used later to constrain the upper and lower uncertainties for the cross section calculations. The NLD of the odd-odd $^{180,182}$Ta are higher than that of the even-odd $^{181}$Ta, due to one extra unpaired neutron in $^{180,182}$Ta which increases the number of degrees of freedom.

\subsection{$\gamma$-ray strength function}

\noindent{}Assuming that the statistical $\gamma$-ray decays are dominated by dipole transitions the $\gamma$SF is given by \cite{Capote}:

\begin{equation}
f(E_{\gamma}) = \frac{\tilde{\mathcal{T}}(E_{\gamma})}{2\pi E_{\gamma}^{3}}.
\end{equation}

\noindent{}The absolute normalisation parameter $B$ is obtained by constraining the data to $\langle \mathit{\Gamma_{\gamma 0}} \rangle$ for s-wave resonances by \cite{Kopecky1990}:

\begin{equation}
\begin{split}
&\langle\varGamma_{\gamma 0}(S_{n})\rangle=\frac{1}{2\pi\rho(S_{n},I_{T},\pi_{T})} \sum_{I_{f}}\int_{0}^{S_n} \times \\
&B\mathcal{T}(E_\gamma) \rho(S_{n}-E_{\gamma}, I_{f})dE_{\gamma},\\
\end{split}
\label{q9}
\end{equation} 

\noindent{}where $\pi$ is the parity, the subscripts $f$ and $T$ indicate the final levels and target nucleus, respectively. 

The photo absorption cross section, $\sigma(E_{\gamma})$, can be converted to the $\gamma$SF by \cite{Bar1973}:


\begin{equation}
\label{eq11}
f(E_{\gamma}) = \frac{\sigma(E_{\gamma})}{3E_{\gamma}(\pi \hbar c)^{2}}.
\end{equation}

\begin{figure*}[!hbt]
	\centering
	\includegraphics[width=1.04\textwidth]{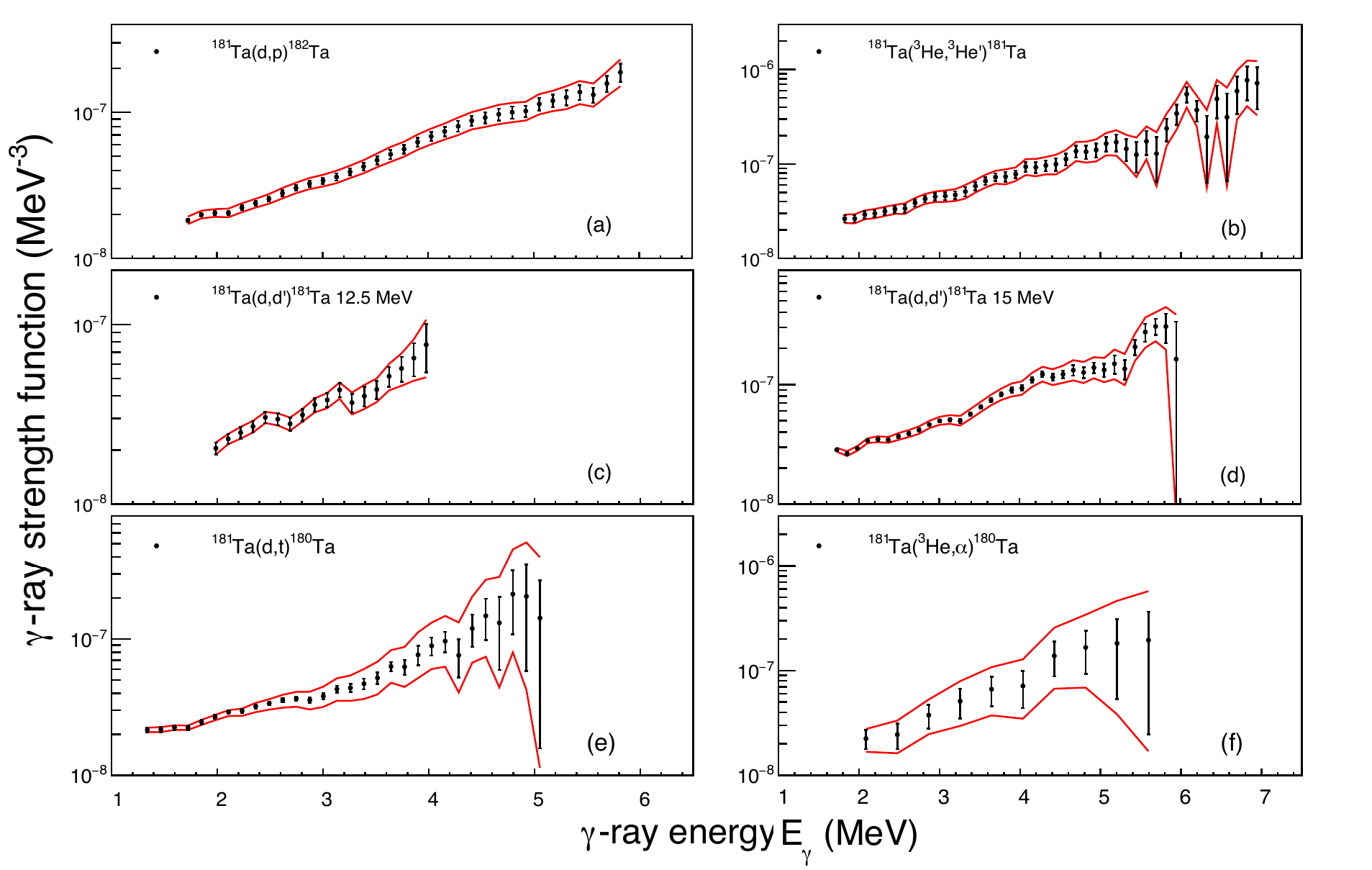}
	\caption{(Color online) The $\gamma$SFs for $^{180,181,182}$Ta for each reaction are shown as black circles with the systematic and statistical uncertainties represented by the error bars. The red lines are the extent of the upper and lower uncertainties which include the experimental uncertainties in $D_{0}$ and $\langle \mathit{\Gamma_{\gamma 0}} \rangle$, in addition to systematic and statistical uncertainties.}
	\label{PSF_indiv}
\end{figure*}

\begin{figure}[h!]
	\includegraphics[width=0.52\textwidth]{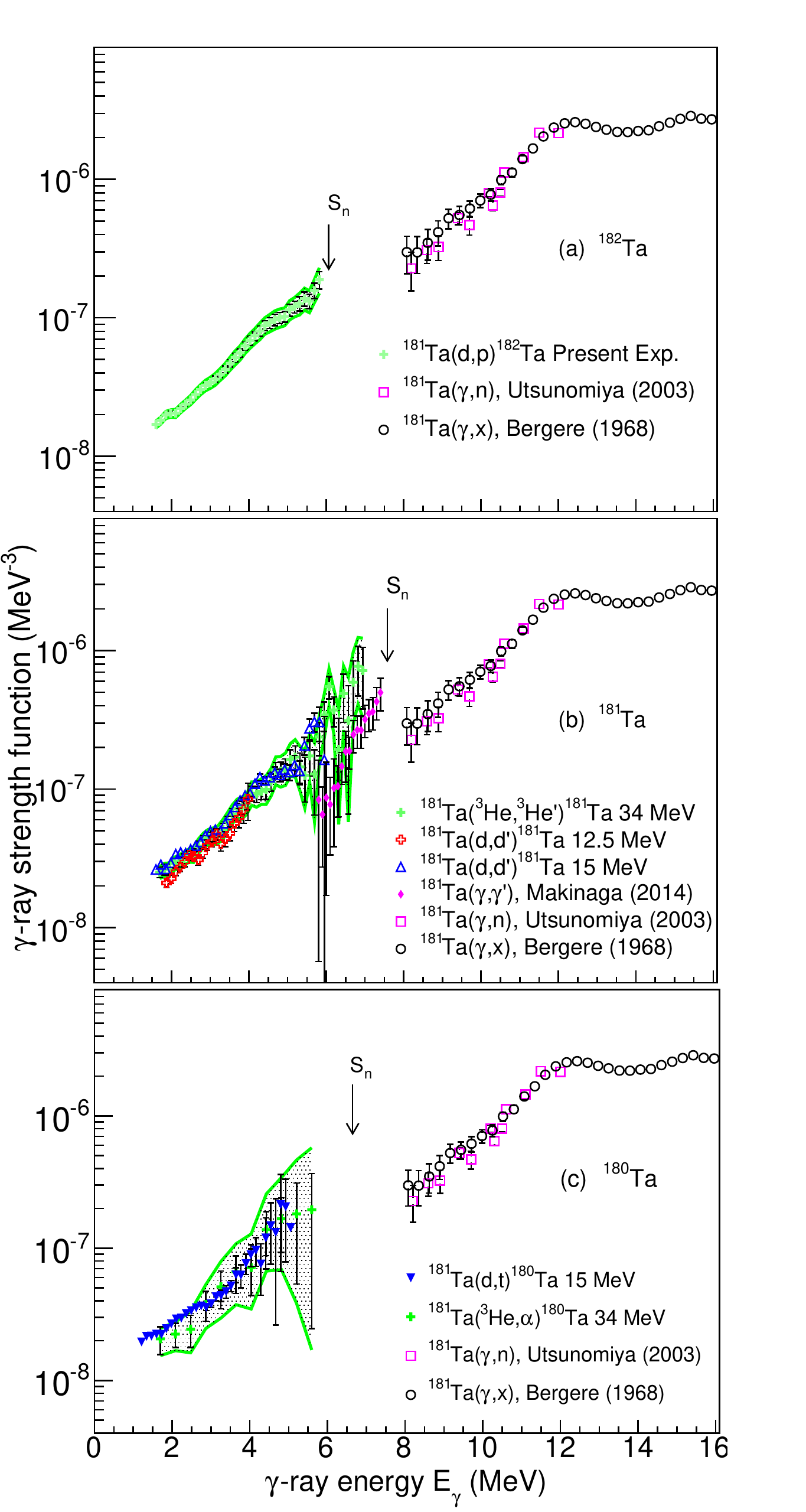}
	\caption{(Color online) The experimental $\gamma$SFs of $^{182}$Ta (a), $^{181}$Ta (b) and $^{180}$Ta (c) from present experiments, are compared to data obtained from $^{181}$Ta($\gamma$,n) \cite{Utsunomiya2003}, $^{181}$Ta($\gamma$,xn) \cite{Bergere1968} and $^{181}$Ta$(\gamma, \gamma)$ \cite{Makinaga2014}. The upper and lower uncertainty bands (green lines) are the combination of statistical, systematic and experimental uncertainties due to D$_{0}$ and $\langle \mathit{\Gamma_{\gamma 0}} \rangle$. Here, they are shown only for the data with the largest uncertainties. }
	\label{GEDRx}
\end{figure}

\noindent{}The extracted $\gamma$SFs for $^{180, 181, 182}$Ta are shown for each reaction individually in Fig. \ref{PSF_indiv}. For $^{182}$Ta (Fig. \ref{PSF_indiv} (a)) the $\gamma$SF is relatively smooth in the measured range with a possible slight enhancement at $\sim$4.5 MeV which has been reported previously in \cite{Igashira1986}. The $\gamma$SFs for $^{181}$Ta exhibit some features which will be discussed in Sec. V. The $\gamma$SF from the $^{180}$Ta($^{3}$He,$\alpha$) reaction had low statistics resulting in larger binning and uncertainties. 
The uncertainties of the $\gamma$SF normalisation introduced by $D_{0}$ and $\langle \mathit{\Gamma_{\gamma 0}} \rangle$ from Refs. \cite{Capote, Mughabghab2006} were considered by separately extracting upper and lower NLDs and $\gamma$SFs for the experimental data, using $D_{0} = D_{0} \mp \delta D_{0}$ and $ \langle \mathit{\Gamma_{\gamma 0}} \rangle \; =\; \langle \mathit{\Gamma_{\gamma 0}} \rangle \pm \langle \delta \mathit{\Gamma_{\gamma 0}} \rangle$ with the CT+FG1. This produces upper and lower error bands. The parameters used to normalise the $\gamma$SFs and NLDs are listed in Tab. \ref{spama}. All $\gamma$SFs for each nucleus are plotted together, with data obtained from $^{181}$Ta($\gamma$,n) \cite{Utsunomiya2003}, $^{181}$Ta($\gamma$,xn) \cite{Bergere1968} and $^{181}$Ta$(\gamma, \gamma)$ \cite{Makinaga2014}, in Fig. \ref{GEDRx}. The $\gamma$SFs for the same nucleus obtained from different reactions are quite similar and agree within the uncertainties.
 
\begin{figure}[h!]
	\centering
	\includegraphics[width=0.53\textwidth]{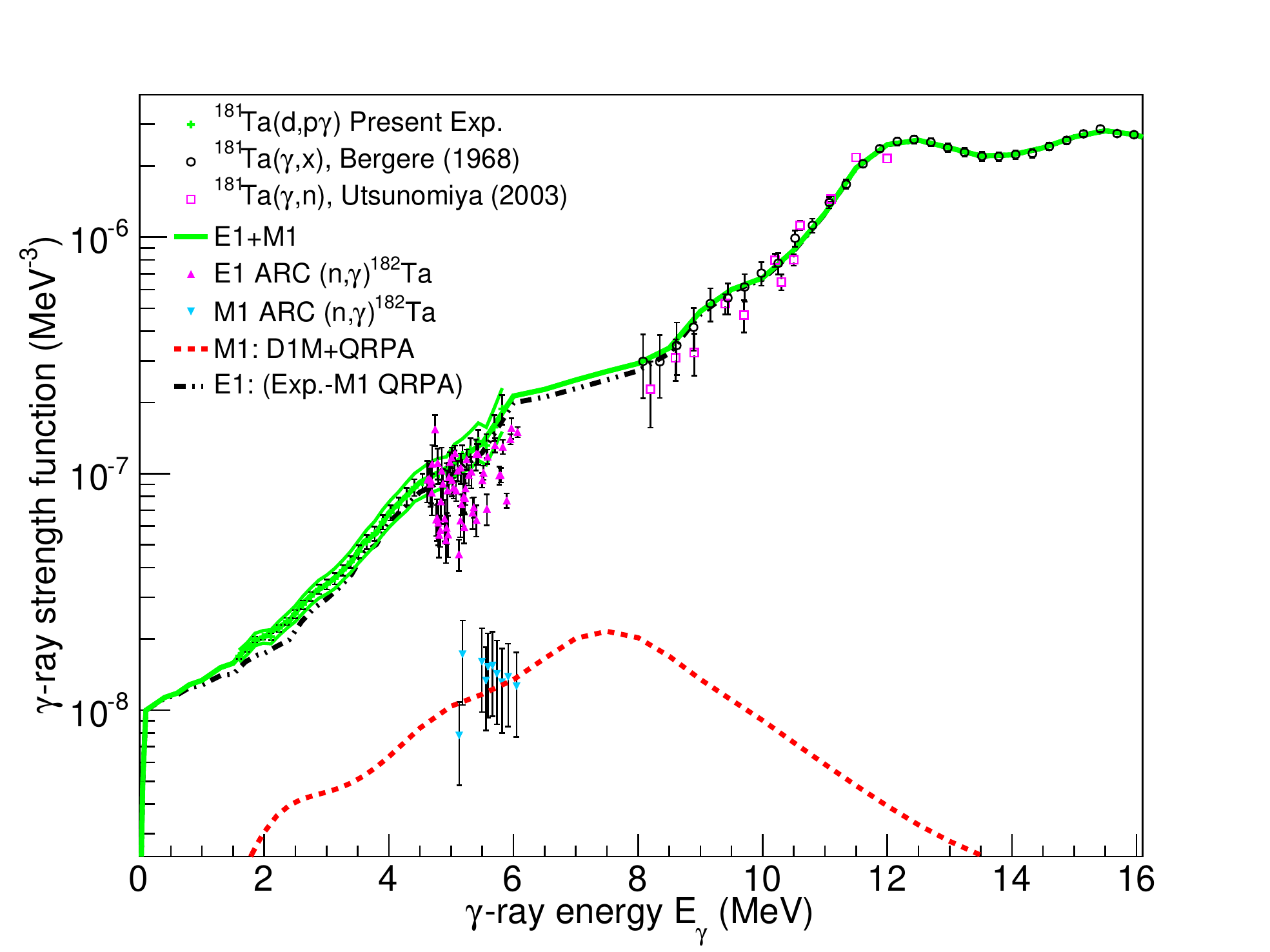}
	\caption{(Color online) The experimental $\gamma$SFs have contributions from E1 and M1 transitions and need to be disentangled. The disentangled E1 and M1 contributions for $^{182}$Ta are compared to ARC data from Ref. \cite{Kopecky2017}.}
	\label{E1M1contributions}
\end{figure}

The experimental $\gamma$SF has contributions from E1 and M1 transitions, and therefore has to be disentangled. This is achieved by subtracting the $M1$ D1M-QRPA strength \cite{Goriely2016, Goriely2018} (Quasi-Particle Random Phase Approximation based on the Gogny D1M interaction) from the experimental E1+M1 $\gamma$SF as shown in Fig. \ref{E1M1contributions}. The disentangled E1 and M1 contributions agree well with average reaction capture (ARC) data from Ref. \cite{Kopecky2017}. The same procedure was applied in the analysis of $^{91,92}$Zr isotopes \cite{Guttormsen2017}. This disentanglement was performed for the experimental strengths from each data set individually.

\section{\label{sec:level4.5}$^{181}$T\MakeLowercase{a}(\MakeLowercase{n},$\gamma$) cross sections}

\noindent{}The E1 and M1 strengths plus the $^{181}$Ta NLDs are used as input in TALYS. The experimental $\gamma$SF span the energy region $E_{\gamma}^{min}$ $\lesssim E_{\gamma}^{exp} \lesssim S_{n}$. The data was extrapolated for $E_{\gamma}^{min} \rightarrow 0$ and $E_{\gamma}^{exp} \rightarrow S_{n}$ to reproduce the experimental $\langle \mathit{\Gamma_{\gamma 0}} \rangle$ values within $<5\%$. Here $E_{\gamma}^{exp}$ is the present experimental data. A linear fit was used to extrapolate the data between the $\gamma$SF and the Giant Electric Dipole Resonance (GEDR) data. 

Whenever possible it is prudent to benchmark existing (n,$\gamma$) cross sections to those that can be obtained using experimental NLDs and $\gamma$SFs. The $^{181}$Ta(n,$\gamma$) cross sections were calculated using the nuclear reactions code TALYS. The key ingredients in the calculations of these (n,$\gamma$) cross sections using the Hauser Feshbach (HF) approach are: the nuclear structure properties (i.e., masses, deformation, $E_{x}$, $J^{\pi}$, etc), NLD, $\gamma$SF and optical model potentials. The global neutron optical potential of \cite{Koning2003} was used for all nuclei in discussion. The Hofmann-Richert-Tepel-Weidenm\"{u}ller-model (HRTW) \cite{Hofmann1975} for width fluctuation corrections in the compound nucleus calculation was used.
\begin{table*}[hbt]
	\caption{The $\gamma$SF and NLD normalisation parameters: resonance spacing $D_{0}$, average radiative width $\langle \mathit{\Gamma_{\gamma 0}} \rangle$, spin-cutoff parameter $\sigma$, level density at the neutron separation energy $\rho (S_n)$ from the CT+FG1, level density parameter $a$, back-shifted energy $E_1$, and the neutron separation energy $S_n$.}
	\setlength{\tabcolsep}{7.0pt}
	\begin{tabular}{|c|c|c|c|c|c|c|c|}
		\hline
		Nucleus & $D_{0}$ (eV) & $\langle \mathit{\Gamma_{\gamma 0}} \rangle$ (meV) & $\sigma (S_n) ^{b}$ & $\rho (S_{n})^{c}$ (10$^{6}$ MeV$^{-1})$ & $a$ (MeV$ ^{-1}$) & $E_{1}$ (MeV)& $S_{n}$ (MeV)\\
		\hline
		$^{180}$Ta & 0.80 $\pm$ 0.24$^{d}$ & 62.0 $\pm$ 5.8$^{d}$ & 4.93 $\pm$ 0.49 & 10.67 $\pm$ 3.50 & 17.57 & -1.09 & 6.65\\
		$^{181}$Ta & 1.11 $\pm$ 0.11$^{a}$ & 51.0 $\pm$ 1.6$^{a}$ & 4.96 $\pm$ 0.50 & 14.58 $\pm$ 2.80 & 17.53 & -0.37 & 7.58\\
		$^{182}$Ta & 4.18 $\pm$ 0.15$^{a}$ & 59.0 $\pm$ 1.8$^{a}$ & 4.88 $\pm$ 0.49 & 2.02 $\pm$ 0.28 & 17.44 & -1.04 & 6.06\\
		\hline
	\end{tabular} 
	\\
	$^{a}$ Average value from \cite{Capote} and \cite{Mughabghab2006}.\\
	$^{b}$ Calculated using Eq. \ref{spin_cut_eqn}\\
	$^{c}$ See text for details.\\
	$^{d}$ No experimental values of $^{180}$Ta are available. See text on how the normalisation parameters were obtained.\\
	\label{spama}
\end{table*} \begin{figure*}
\centering
\includegraphics[width=0.84\textwidth]{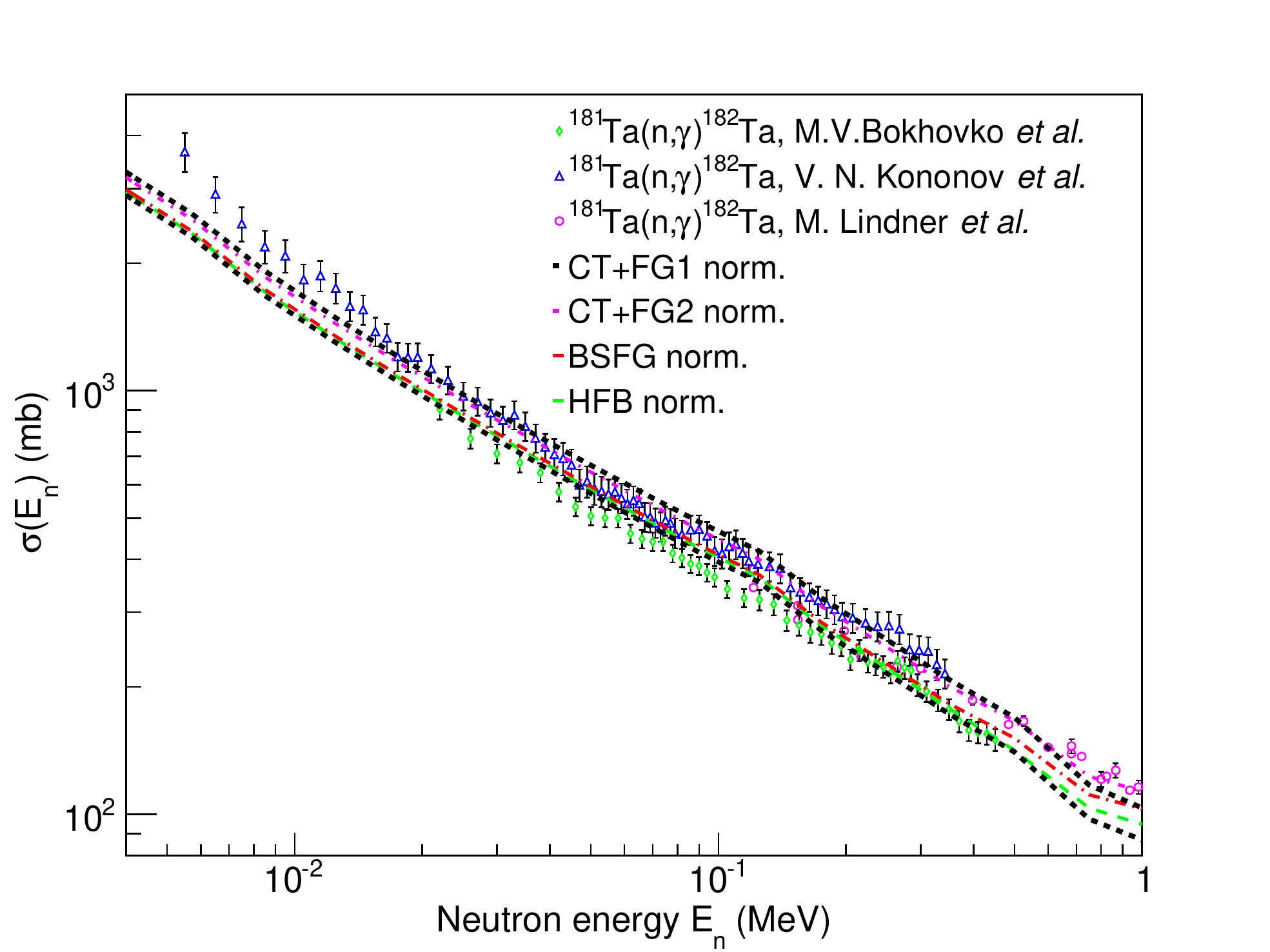}
\caption{(Color online) The present $^{181}$Ta$(n,\gamma)$ cross section (bands) obtained from the $^{181}$Ta(d,p)$^{182}$Ta reaction, compared to previous measurements \cite{Bokhovko1991,Kononov1977,Lindner1976}. The upper and lower band is indicated by the black-dotted line obtained using the CT+FG1 model.}
\label{XS_one}
\end{figure*} \begin{figure}
\centering
\includegraphics[width=0.53\textwidth]{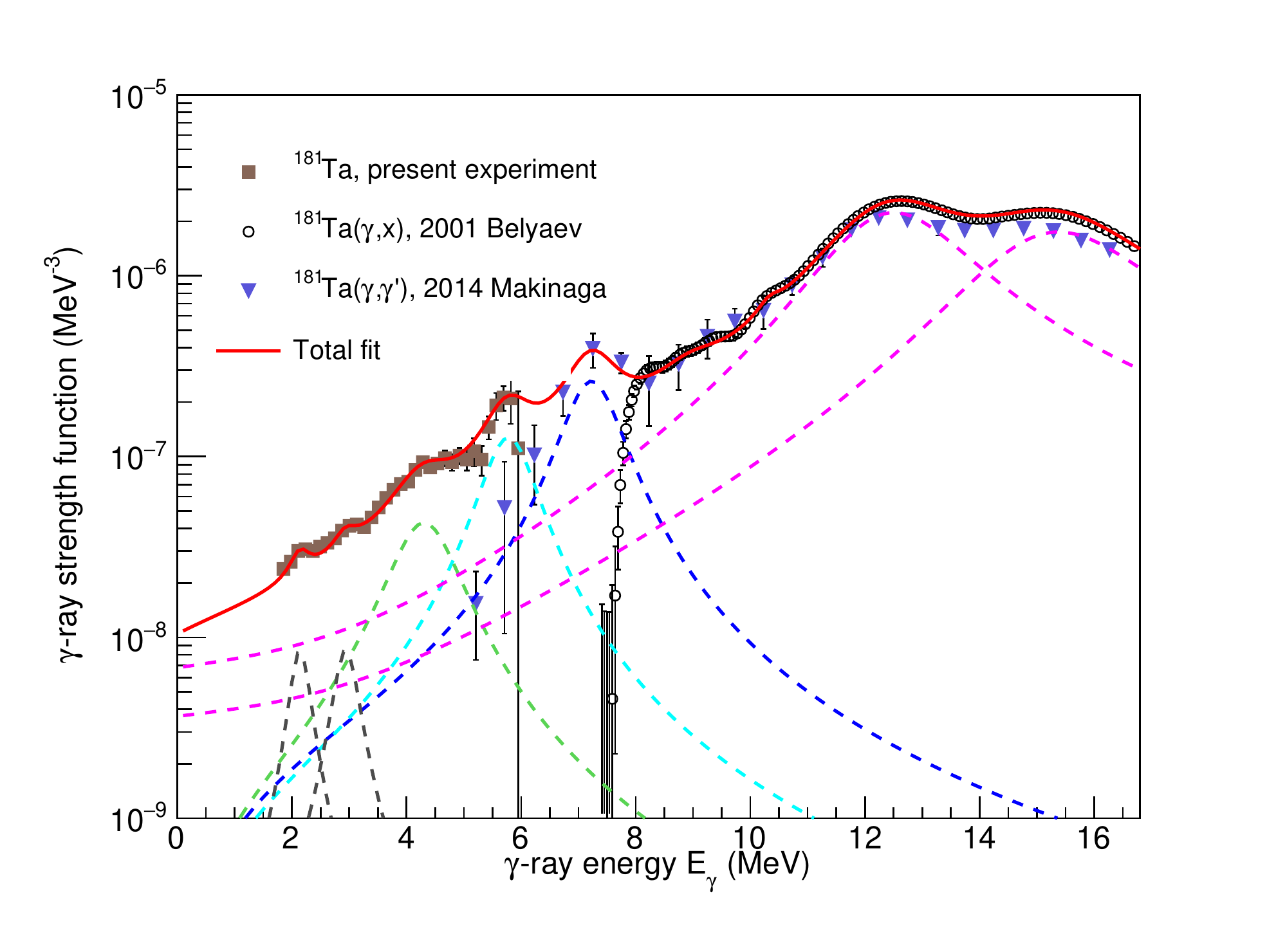}
\caption{(Color online) $^{181}$Ta data from the 15 MeV $^{181}$Ta(d,d')$^{181}$Ta, $^{181}$Ta($\gamma$,$\gamma'$) \cite{Makinaga2014}, and $^{181}$Ta($\gamma,X$) \cite{Belyaev2001} reactions. Various resonances were identified (see text for details) and contribute to the total fit (red line) that best matches the experimental data. } 
\label{reson}
\end{figure}
The $^{181}$Ta(n,$\gamma$) cross sections have been extensively measured in time-of-flight \cite{Bokhovko1991,Kononov1977} and activation \cite{Lindner1976} measurements. It is interesting to compared these cross sections with those obtained from this work. The $^{181}$Ta(n,$\gamma$)$^{182}$Ta cross sections, $\sigma(E_{n})$, as a function of incident neutron energies for 0.004 keV to 1 MeV, taking into account the uncertainties affecting the $\gamma$SFs and the NLDs, have been calculated and are shown in Fig. \ref{XS_one}. The cross sections obtained from the different normalizations yield very similar results. The $^{181}$Ta(n,$\gamma$)$^{182}$Ta cross sections exhibit a slight divergence below $10^{-2}$ MeV, but good agreement above $10^{-2}$ MeV with each other and with previous measurements. Similar results have been observed in Ref. \cite{Kheswa2017}, where different normalisation models and spin distributions were explored in detail, yielding the same results. The agreement further validates that experimental NLDs and $\gamma$SFs can be used to obtain (n,$\gamma$) cross sections indirectly, and gives confidence in this technique to determine reliable (n,$\gamma$) cross sections for which direct measurement techniques are not currently viable e.g. Refs. \cite{Spyrou2014,Kheswa2015}.

\section{\label{sec:level4}Scissors resonance}

\noindent{}The SR is a collective excitation mode dominated by single-particle events usually found at $E_{\gamma} = 66\delta A^{-1/3}$, where $\delta$ is the quadrupole deformation parameter and $A$ is the nuclear mass \cite{Richter1995}. On a macroscopic level the SR may be described by the oscillation of the proton and neutron distributions against each other, similar to scissor blades. On a microscopic level the SR originates from transitions between Nilsson orbits of $\Delta \Omega = \pm 1$ with the same spherical $j$ component. The quantum number $\Omega$ is the projection of the total angular momentum onto the symmetry axis of the nucleus. 

A splitting of the SR may be interpreted by means of $\gamma$ deformation along the three axes \cite{Iudice1985}:

\begin{equation}
\label{SR_eqn}
\begin{split}
\omega_{1} = (\cos \gamma + \left(\frac{1}{\sqrt{3}}\right)\sin \gamma) \omega_{M1},\\
\omega_{2} = (\cos \gamma - \left(\frac{1}{\sqrt{3}}\right)\sin\gamma)\omega_{M1},\\
\omega_{3} = \omega_{M1}\frac{2}{\sqrt{3}}\sin\gamma, 
\end{split}
\end{equation}

\noindent{}where $\omega_1$, $\omega_2$, and $\omega_3$ are the centroid energies of the individual SR components and $\omega_{M1}$ is the energy resonance centroid. Along the third axis, $\omega_{3}$ is located at low energies which is typically not within experimental reach of the Oslo Method. The splitting of the SR of the two higher-lying components can be calculated by \cite{Iudice1985}:

\begin{equation}
\label{SR_split}
\Delta \omega = \omega_{1} - \omega_{2} = \omega_{M1}\frac{2}{\sqrt{3}}\sin\gamma.
\end{equation}

\noindent{}For axially symmetric nuclei ($\gamma$=0) the $\omega_{3}$ component is absent and the $\omega_{1}$ and $\omega_{2}$ components are degenerate. 

Cross sections from ($\gamma$,$\gamma'$) and ($\gamma,x$) reactions \cite{Makinaga2014, Belyaev2001} were converted to $\gamma$SF data with Eq. \ref{eq11}. The resonances of $^{181}$Ta for $E_{\gamma} <$ 9 MeV were fitted with standard Lorentzian functions, while for the components of the GEDR (purple dashed lines), the enhanced generalised Lorentzian functions were used, as shown in Fig. \ref{reson}. The GEDR parameters were slightly modified from the average values of Refs. \cite{Capote, Mughabghab2006} to better match the experimental data. From $(\gamma,\gamma')$ data the enhanced $\gamma$SF, for 6 MeV $<E_{\gamma}<$ 8 MeV (dark-blue dashed line in Fig. \ref{reson}) was suggested to be due to the E1 pygmy resonance \cite{Makinaga2014}. A slight change in the gradient at around 4.5 MeV was noted for $^{182}$Ta in \cite{Igashira1986}, and this feature is also visible in our data and assumed to be a resonance at $\sim$ 4.3 MeV (green dashed line in Fig. \ref{reson}). An additional unknown resonance at 5.8 MeV (light-blue dashed line in Fig. \ref{reson}) was added so that the total fit matches the experimental data. The resonance parameters used for the fits in Fig. \ref{reson} are shown in table \ref{res_par}.

\begin{table}[!hbt]
\caption{The resonance centroid $\omega$, amplitude $\sigma$ and half width at half maximum $\Gamma$ used to fit the $\gamma$SF resonances. Enhanced generalised Lorentzian functions were used to fit the GEDR and standard Lorentzian functions were used for the other resonances.}
\label{res_par}
\setlength{\tabcolsep}{15.2pt}
\begin{tabular}{|c|c|c|c|c|}
\hline
 $\omega$ (MeV) & $\sigma$ (mb) & $\Gamma$ (MeV) \\
\hline
2.2 & 0.2 & 0.4 \\
 2.9 & 0.3 & 0.5 \\
 4.4 & 2.3 & 1.3 \\
 5.8 & 8.5 & 1.0\\
 7.3 & 21.8 & 1.1 \\
 12.7 & 340 & 2.8\\
 15.6 & 320 & 3.6\\

\hline
\end{tabular} 
\\
%
\end{table}

The $\gamma$SF of $^{181}$Ta exhibits weak features at 2 MeV $<$ E$_{\gamma}$ $<$ 3.5 MeV (black dashed lines in Fig. \ref{reson}, which are found in the typical energy range for the SR \cite{Kris2010}. From this work the distinction between M1 and E1 is not possible but the assignment to the SR and its location in $^{181}$Ta is corroborated by previous measurements \cite{Wolpert1998, Angell2016}.

The SR splits into two peaks, at $E_{\gamma} =$ 2.16 $\pm$ 0.04 MeV and $E_{\gamma} =$ 2.91 $\pm$ 0.05 MeV, which is consistent with the fragmentation observed in Ref. \cite{Wolpert1998}. The average splitting of the SR peaks in $^{181}$Ta is $\Delta \omega =$ 0.75 $\pm$ 0.06 MeV. Using Eq. \ref{SR_split} a $\gamma$ deformation of 14.9$^{\circ}$ $\pm$ 1.8$^{\circ}$ is calculated. No additional strength is observed for $^{180}$Ta or $^{182}$Ta in the energy region of the SR.

\begin{figure}[!t]
	\centering
	\includegraphics[width=0.49\textwidth]{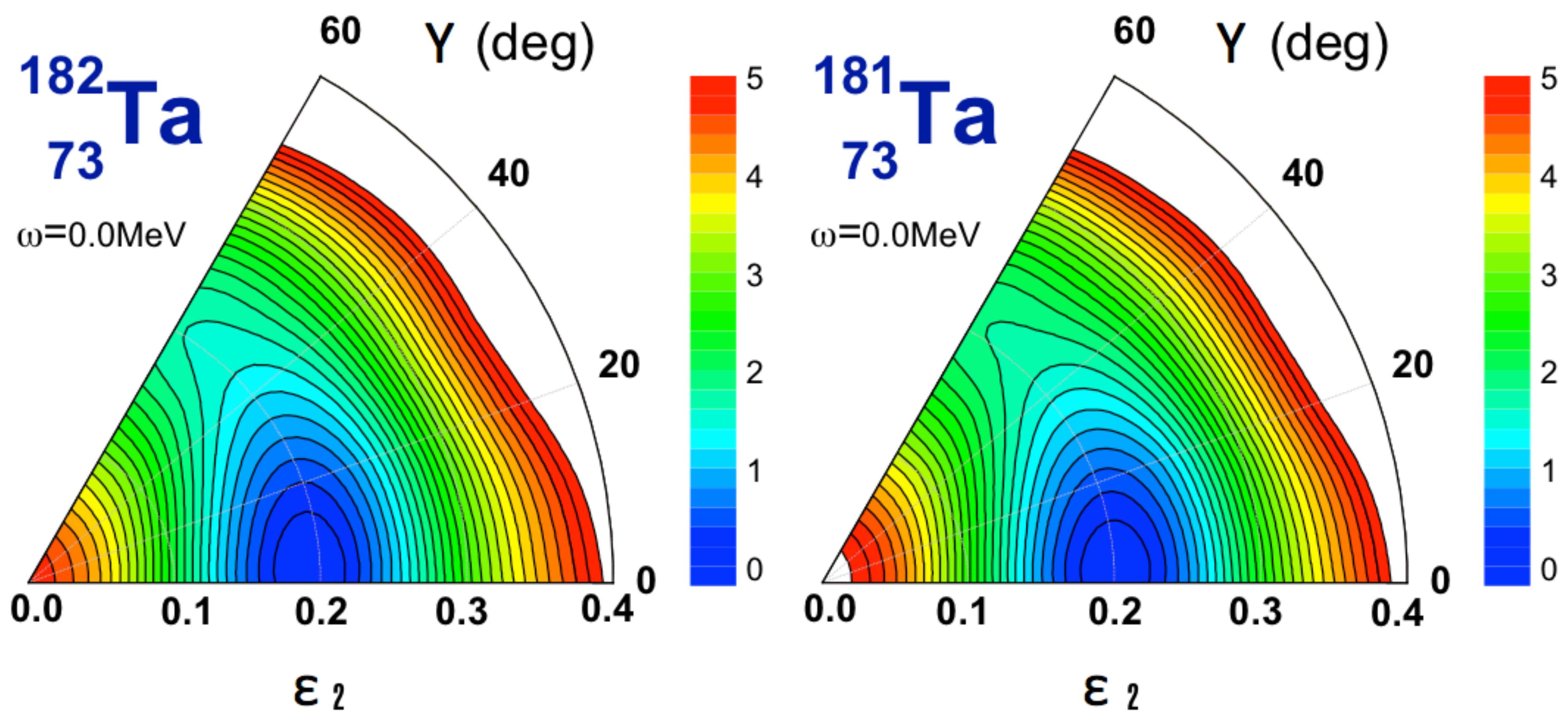}
	\caption{(Color online) Potential energy surface calculations with the Cranking Nilsson model plus Shell correction method for the ground states of $^{181,182}$Ta, see text for details.} 
	\label{gamma_def}
\end{figure}

Potential energy surface calculations for $^{181,182}$Ta were performed with the Cranking Nilsson model plus Shell correction method \cite{Frauendorf2000,Dimitrov2000,Xu2018} with pairing-gap values adopted from Ref. \cite{1997Moller} and are shown in Fig. \ref{gamma_def}. From these it is apparent that the ground-state configuration in $^{181}$Ta and $^{182}$Ta exhibit a $\gamma$-axis minimum, between 0$^{\circ}$-15$^{\circ}$ and a deformation parameter of $\epsilon_2 \approx$ 0.2. The deformation parameters $\delta$ and $\epsilon_{2}$ are the same to first order. From this, $^{181,182}$Ta exhibit some softness towards $\gamma$ in the form of $\gamma$-vibrations and collectively prolate which is in agreement with $\gamma = 14.9^{\circ}$ $\pm$ 1.8$^{\circ}$ extracted from the splitting of the SR. This $\gamma$ deformation is also in agreement with those predicted in Refs. \cite{Hilaire2007,Delaroche2010}.

The neutron capture $\gamma$-ray spectra \cite{Igashira1986} of the odd-odd nuclei $^{142}$Pr, $^{160}$Tb, $^{166}$Ho, $^{176}$Lu, $^{182}$Ta, and $^{198}$Au are particularly interesting and can shed light on the above results. The large deformation of $\epsilon_2 \sim$ 0.32 \cite{Hilaire2007} in $^{160}$Tb appears to produce a relatively localised strength at $E_{\gamma}$ = 2.5 MeV despite the two odd nucleons. Fragmentation increases for $^{166}$Ho and $^{176}$Lu as deformation is somewhat reduced to $\epsilon_2 \sim$ 0.30 \cite{Hilaire2007}. For $^{142}$Pr, $^{182}$Ta, and $^{198}$Au deformation is further reduced and may explain why the resonance is not identifiable. This is consistent with the proportionality of $B(M1)$ with the square of deformation \cite{Ziegler1990}. While higher detection sensitivity \cite{Wolpert1998} reveals the presence, albeit fragmented, of the SR in $^{181}$Ta, the additional odd neutron and a slightly reduced deformation is sufficient to fragment the SR strength to a level that it is not observable in $^{180}$Ta and $^{182}$Ta.

Low-lying excitations of $^{181}$Ta were investigated using NRF experiments \cite{Angell2016, Wolpert1998}. It was suggested that the SR was rather weak and splits into two parts. From our work, it can be concluded that a weak SR is observed with split centroids located at 2.16 MeV $\pm$ 0.04 MeV and 2.91 MeV $\pm$ 0.05 MeV, in agreement with NRF measurements \cite{Angell2016, Wolpert1998}. The case of $^{182}$Ta is similar to that of $^{197,198}$Au \cite{Giacoppo2015} where no SR is observed.

The current results support nuclear triaxiality as the likely mechanism of SR splitting in $^{181}$Ta however there are alternative explanations. The SR splitting was proposed from microscopic calculations \cite{Balbutsev2013}, which were able to explain the observed splitting in the actinide region \cite{Tornyi2014,Guttormsen2012, Guttormsen2014,Laplace2016}, where the triaxiality argument does not hold due to a mismatch of $B(M1)_{\omega2}/B(M1)_{\omega1}$, from the $B(M1)$ values of the individual SR components, and from the extracted $\gamma$ deformation \cite{Guttormsen2012}. In these calculations the SR mode of protons oscillating against neutrons is accompanied by a lower-energy nuclear spin scissors mode where spin-up nucleons oscillate against spin-down nucleons.  Despite systematic axially deformed QRPA calculations \cite{Goriely2016, Goriely2018}, the evolution of the SR across the nuclear chart is still not fully understood. For a complete understanding of the interplay of the SR with other nuclear structure properties, such as the coupling to unpaired nucleons and its dependence on nuclear shape, the persistence of the SR in transitional regions of the nuclear chart has to be investigated further. 

\section{Summary}

\noindent{}The NLDs and $\gamma$SFs of $^{180,181,182}$Ta were measured at the Oslo Cyclotron Laboratory. Six independent data sets from $^{181}$Ta(d,X) and $^{181}$Ta($^{3}$He,X) reactions were analysed with the Oslo Method. The total NLDs at the neutron separation energies and their uncertainties were calculated using three different models, the BSFG, CT+FG (1,2), and HFB plus Combinatorial models. 

The comparison between the $^{181}$Ta(n,$\gamma$) cross-sections calculated with TALYS v1.9 using the measured NLD and $\gamma$SF and the results from direct measurements is satisfying and reinforces the appropriateness of using NLDs and $\gamma$SFs for the determination of neutron capture cross sections. 

The $\gamma$ deformation of 14.9$^{\circ}$ $\pm$ 1.8$^{\circ}$ for $^{181}$Ta was calculated and this $\gamma$ softness, together with the unpaired nucleon, may be an explanation for a significant fragmentation of SR strength. Nuclear triaxiality may be considered as the likely mechanism of the observed SR splitting in $^{181}$Ta, but further experimental work and theoretical guidance on possible observables and specific experimental signatures for the spin-SR mode are desirable. 

\section*{Acknowledgments\protect}

\noindent{}We would like to thank J. C. M{\"u}ller, A. Semchenkov, and J.C. Wikne for providing quality beam. The authors thank A.O. Macchiavelli for insightful discussions. This work was supported by the National Research Foundation of South Africa under grant nos. 100465, 92600, and 92789 and the US Department of Energy under contract no. DE-AC52-07NA27344. A.C.L. acknowledges support from the ERC-STG-2014 under grant agreement No. 637686. The authors gratefully acknowledge funding from the Research Council of Norway, project grant no. 222287 (G.M.T.), 263030 (A.G., V.W.I., F.Z. and S.S.), 213442 and 263030. S.G. acknowledges the support of the FRS-FNRS. This work was performed within the IAEA CRP on ``Updating the Photonuclear data Library and generating a Reference Database for Photon Strength Functions'' (F410 32). M. W. and S. S. acknowledges the support from the IAEA under Research Contract 20454 and 20447, respectively. 

\bibliography{Reference}

\end{document}